# Dataset: Analysis of IFTTT Recipes to Study How Humans Use Internet-of-Things (IoT) Devices

Haoxiang Yu
hxyu@utexas.edu
The University of Texas at Austin
Austin, TX, USA

Jie Hua
mich94hj@utexas.edu
The University of Texas at Austin
Austin, TX, USA

Christine Julien
c.julien@utexas.edu
The University of Texas at Austin
Austin, TX, USA

## ABSTRACT

With the rapid development and usage of Internet-of-Things (IoT) and smart-home devices, researchers continue efforts to improve the "smartness" of those devices to address daily needs in people's lives. Such efforts usually begin with understanding evolving user behaviors on how humans utilize the devices and what they expect in terms of their behavior. However, while research efforts abound, there is a very limited number of datasets that researchers can use to both understand how people use IoT devices and to evaluate algorithms or systems for smart spaces. In this paper, we collect and characterize more than 50,000 *recipes* from the online *If-This-Then-That* (IFTTT) service to understand a seemingly straightforward but complicated question: "What kinds of behaviors do humans expect from their IoT devices?"

## 1 INTRODUCTION

During the past decade, people have increasingly been using IoT devices within their homes and other smart spaces. At the same time, significant commercial and research efforts have sought to simplify the creation of smart spaces, whether to allow automation of integrations of smart devices [4] or to create middleware that eases the programming burden associated with leveraging smart devices within applications [2, 3, 8]. However, a lack of robust datasets that capture everyday users' interactions with devices in their smart spaces slows research. Without a realistic dataset of IoT device interactions in smart homes, the community lacks (1) a clear understanding of the scope of interactions between humans and IoT devices on which to base further development and (2) a mechanism to support large-scale and repeatable evaluation of smart home IoT solutions. This paper seeks to change this situation.

In this dataset, we seek to characterize humans' interactions with IoT devices in their smart spaces by capturing rules, behaviors, and policies that those humans define and deploy to their spaces.

To do so, we rely on the widely used *If-This-Then-That* (IFTTT)[1] service. IFTTT and other RESTful APIs like it (e.g., Zapier[2] and Automate.io[3]) effectively provide templates by which end-users can create mash-ups of their digital services to define automatons that represent their personalized desired outcomes [7]. Generically, the rules in these systems connect *triggers* to *actions* that are taken when the triggers are engaged. For instance, some of the most popular rules on IFTTT automate calendar notifications based on the weather or send time-triggered reminders via common messaging channels. Rather than concerning ourselves with automation rules *in general*, the focus of our dataset is on IFTTT rules (called *recipes*) that connect one or more IoT devices with one or more pieces of automatically collected context information. For instance, a common IFTTT rule instructs a user's smart lawnmower to park itself if heavy rain is predicted in the weather forecast.

Other efforts have also sought to characterize behaviors that users encapsulate in automation rules in IFTTT [6, 9, 10]. While these efforts collected large numbers of rules and characterized them, they are (1) out of date, collected before many current IoT devices were on the market, and (2) focused on rules in general rather than on those designating smart space interactions specifically.

To construct the dataset described in this paper, we collected 50,067 publicly shared IFTTT recipes. We filtered the collected rules based on their popularity, relevance to IoT applications, and ease of automated parsing, resulting in 2,648 rules that are included as part of our final IoT behavior dataset. We further analyzed and characterized the patterns in these rules, as described throughout this paper. We packaged the resulting rules in an easily ingestible format that other researchers and developers can use to understand how humans use IoT devices in their smart spaces and to evaluate emerging IoT smart space systems and algorithms.

## 2 DATA COLLECTION

Figure 1 shows our data collection process. We started by collecting publicly available IFTTT recipes from the IFTTT website[4]. To seed our searches, we used the 10,000 most used English words [5]. Each recipe we collected is associated with a unique identifier, which we used to remove any duplicate recipes. After this step (step 1 in Figure 1), we had discovered 50,067 recipes. These recipes are all included in the raw data portion of our final dataset.

Each IFTTT recipe is associated with a set of features that describe the recipe. Several of these are not useful in the context of our dataset, and we removed them to reduce the size of the dataset. Table 1 shows an example IFTTT rule with some of its most informative features included.

[1]https://ifttt.com
[2]https://zapier.com
[3]https://automate.io
[4]Recipes were collected on August 20, 2021 from https://ifttt.com/search/query/.

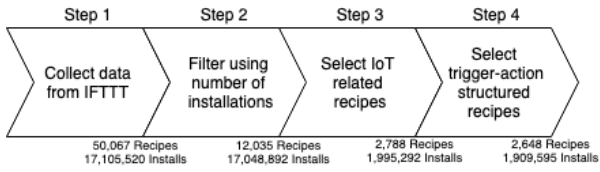

**Figure 1: Data collection steps**

**Table 1: Selected features for an example IFTTT recipe**

| Feature | Example contents |
|---|---|
| name | Automatically turn your lights on at sunset |
| id | PVkgiLYy |
| services | 'weather', 'hue' |
| service names | 'Weather Underground', 'Philips Hue' |
| description | Never be left in the dark. Whenever the sun starts to set, your Philips Hue bulbs will automatically turn on. |
| by service owner | TRUE |
| pro features | FALSE |
| installs | 144,011 |

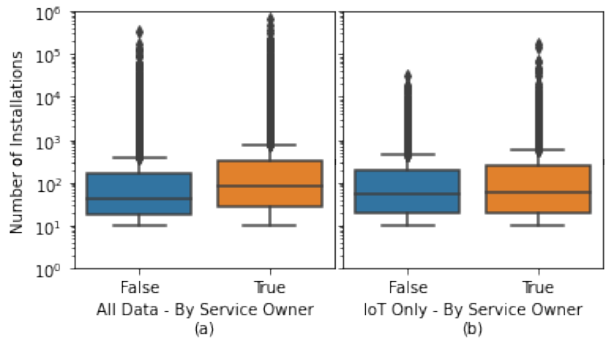

**Figure 2: Installation distributions (Log scale)**

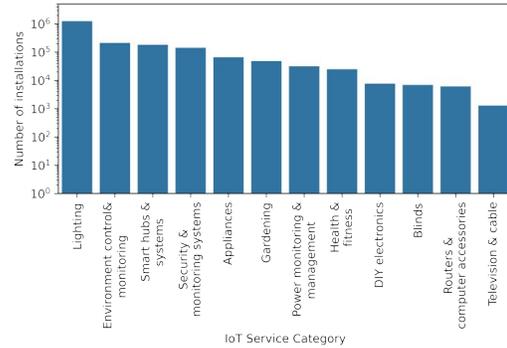

**Figure 3: Installations for recipes involving IoT devices (Log scale)**

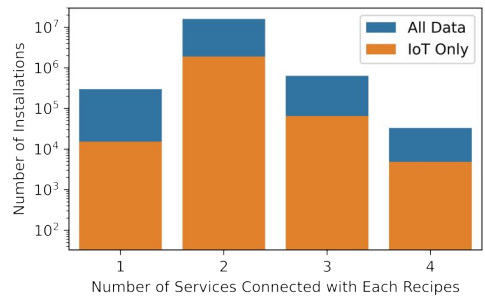

**Figure 4: Frequencies for the number of services per recipe (Log scale)**

Because our goal is to identify common patterns in human behavior, we next honed in on rules that are more frequently applied by users, by filtering based on the number of installations of each IFTTT recipe (step 2 in Figure 1). In particular, we observed that 53% of the published recipes are used by only 0 or 1 user; only 24% of the recipes have 10 or more users. Therefore, we filtered the set of rules to include only those with 10 or more users, resulting in a set of 12,036 *popular rules*. Nonetheless, even after filtering out rules with few instances of installation, the distribution of popularity of the remaining rules remains uneven, as shown in Figure 2(a) The figure also shows that there is a slight increase in the number of installations for IFTTT recipes that are provided by service owners (e.g., like the rule in Table 1, which is defined by Philips, the manufacturer of the Hue series of smart lights), though the increase is not significant.

Next, many of the rules in the dataset provided service automation that has nothing to do with smart environments (e.g., "Every morning at 7 am, send a Slack message with the first meeting of the day from Google Calendar"). Since our goal is to assemble a dataset that captures humans' interactions with IoT devices specifically, we collected category information from the IFTTT services list[5]; mapped the 658 different services that were used in our *popular rules* to the 47 IFTTT service categories; and then manually select categories that are related to IoT devices (step 3 in Figure 1). These categories are shown across the x-axis in Figure 3.

This procedure leaves us with 2,788 *popular IoT rules.* Figure 2(b) shows the distribution of installations for only the IoT-relevant IFTTT recipes in the dataset. As the figure shows, IoT-relevant rules are overall less frequently used than the average of the dataset, and the slight increase for rules defined by the service owner almost disappears. Figure 3 shows the distribution of installation by IoT category; clearly, IFTTT recipes that control the ambient lighting are by far the most popular among all of the IoT device categories.

Finally, we filtered the dataset to include rules that conjoin only a single trigger with a single action. Figure 4 characterizes both the original dataset and the filtered *popular IoT rules* dataset in terms of the number of services involved in each IFTTT recipeÅs the figure shows, it is most common for a recipe to connect just two services, similar to the example in Table 1. However, some IFTTT recipes reference only a single service (e.g., "Play a spoken notification when your battery is low"), while others reference three, four, or even five services. In the latter category, these rules may connect one or more triggers with one or more actions to be taken on that trigger (e.g., "Turn on my Hue lights when I arrive home unless it's still light outside"). For our final dataset, we selected only rules with a single trigger and a single action to better ensure that we capture intentional behaviors that users specify for their IoT environments. Ultimately, our final filtered dataset consists of 2648 rules for human-specified IoT interactions.

---
[5]https://ifttt.com/services

Table 2: Distribution for the binomial information

|  | All Rules | | IoT Rules | |
|---|---|---|---|---|
|  | TRUE | FALSE | TRUE | FALSE |
| *Made by Service Owner* | 29.85% | 70.15% | 43.54% | 56.46% |
| *Pro Features* | 2.43% | 97.56% | 3.48% | 96.52% |
| *Requests Mobile App* | 24.19% | 75.81% | 15.93% | 84.07% |

## 3 DATA ANALYSIS

In this section, we provide some high-level observations about the collected dataset, as well as some concrete examples of various rules contained within the dataset.

First, Table 2 shows three key features of IFTTT rules: (1) whether the rule was defined by the service owner (e.g., like the Philips Hue defined rule in Table 1); (2) whether the rule makes use of IFTTT "pro"-level features [6], and (3) whether the rule depends on a mobile application. Each of these is represented by a boolean flag in the rule's entry in the dataset; we show the distribution of TRUE and FALSE flags for all 50,067 rules in the dataset and for the 2,648 IoT rules in our final filtered dataset. As the table shows, IoT rules are significantly more likely to be defined by the service owner, slightly more likely to require pro features of the IFTTT platform, and significantly less likely to require a mobile app.

Next, because we seek to understand how a user's triggers are associated with their actions in an IoT smart space, we manually cataloged the 2,648 rules based on two dimensions: the type of action (i.e., roughly the user's goal) and the type of the trigger.

The types of actions include the following:

- **Ambient Temperature** rules adjust the "feels-like" temperature (e.g., adjust A/C, turn on a fan, open a window).
- **Ambient Luminance** rules adjust the lighting (e.g., turn on/off a light, adjust the light level, open the blinds).
- **Security** rules are those that involve any smart security device (e.g., alarm system, camera, lock, garage door).
- **Alert User** rules have some mechanism to capture the user's attention (e.g., blink the light, make a voice announcement, change displayed information).
- **Energy Saving** rules are those that intend to save energy (e.g., turn off the A/C when the user leaves, turn off lights at night).
- **Ambient Atmosphere** rules change the atmosphere for entertainment purposes (e.g., change the color of the lighting).
- **Robot Control** rules control the actions of some robot (e.g., robot vacuum or lawnmower)
- **Control Hub** rules relate to hub devices that can in turn control multiple other devices but for which the ultimate goal is not obvious.

- **Gardening** rules are those that impact the outdoors but do not involve robots or ambient lighting (e.g., control watering).
- **Outlet Control** rules control a smart electrical outlet, without any additional information about the connected device.
- **Other Appliances** rules reference home appliances not listed above (e.g., coffee maker, microwave, dishwasher).

Any rules that did not explicitly fall into one of these categories is labeled **Other**, e.g., controlling game consoles, feeding pets, etc. For the second dimension, we defined six trigger types:

- **Explicit Control (Voice)** rules are triggered by voice commands. These rules commonly attempt to connect to another smart home device (e.g., control thermostat via voice speaker).
- **Explicit Control (Button)** rules are triggered by a virtual button on a phone or a physical button in the space.
- **Spatial Trigger** rules have actions based on the user's location whether detected via GPS or a Wi-Fi connection.
- **Time Trigger** rules run at some specific time or date.
- **Weather Trigger** are triggered by current or forecast weather.

In Figure 5, we show the distribution of installations of rules in our final dataset that fall into each pair of trigger and action combinations. In the figure, the percentages in each cell indicate the fraction of rules for that action category that use the corresponding trigger. For instance, 37.2% of the ambient temperature action rules rely on explicit voice triggers. The numbers under each action category indicate the number of unique rules in that category and the number of installations of rules in that category. For instance, there are 359 unique ambient temperature rules, which have, in total 206,466 installations. Finally, for each combination of action category and trigger type, we give one or two example rules from the dataset in Table 3, where the numbers in the leftmost column in the table correspond to the numbers in the cells in Figure 5.

## 4 DATASET ACCESS

The dataset can be downloaded at [1] [7] This dataset should be used for academic non-commercial purposes only. Any other usage is prohibited according to the IFTTT Terms of Service

## 5 CONCLUSIONS

The presented dataset provides insights into how humans use Internet-of-Things (IoT) devices. This information can help research teams build large-scale evaluations of it or similar systems. As examples of specific use cases, since *conflicts* among policies in the use of IoT devices is a commonly research challenge, these example rules could be used to create policies that are realistic and yet exhibit conflicting uses of a set of IoT devices. Alternatively researchers of new IoT devices may use the information contained in the dataset to develop new devices or applications that are ever more tailored to the specific automations that real users desire. More generally, the dataset could be an input to evaluations of wide ranging systems, middleware, and applications that are constructed to support human interactions in the IoT.

---

[6] https://ifttt.com/pro

[7] Email hxyu@utexas.edu to request data access

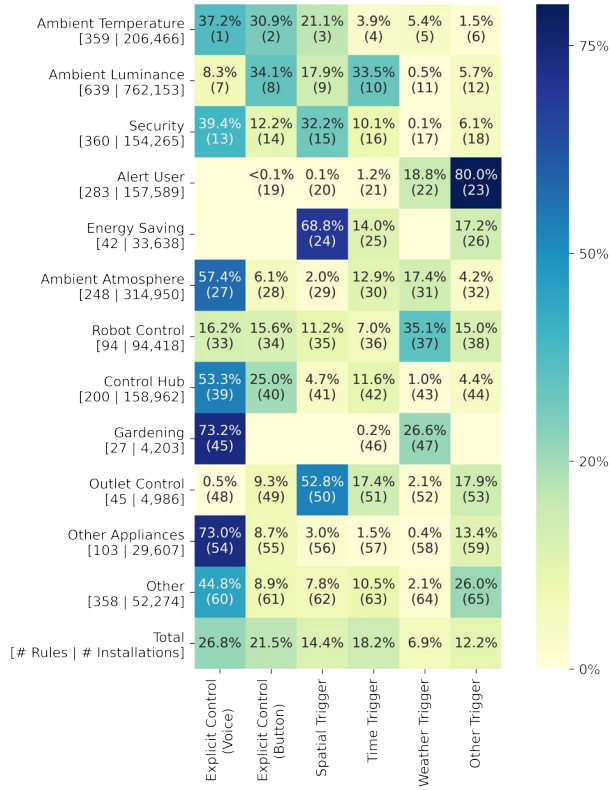

**Figure 5: Percentage of each trigger with each category. Example for each cell is listed in Table 3**

## ACKNOWLEDGMENTS

This work was funded in part by the National Science Foundation under grants CNS-1813263 and CNS-1909221. Any opinions, findings, conclusions, or recommendations expressed are those of the authors and do not necessarily reflect the views of the NSF.

## Table 3: Example Recipes for Each Cell in Figure 5

| Cell | Example Rules | Trigger Type |
|---|---|---|
| *Ambient Temperature* | | |
| (1) | "Tell Amazon Alexa to set Nest temperature" or "Use Google Assistant to run Nest fan" | Voice |
| (2) | "Set your comfort profile with the push of a button" or "Boost heating for 1 hour at 22°C in one tap" | Button |
| (3) | "Turn on your Nest Thermostat as you arrive home" or "If you exit an area then turn your heating off" | Spatial |
| (4) | "At sunrise turn your fan on for 15 minutes" or "At midnight, set heating back to auto." | Time |
| (5) | "Automatically turn off AC if it's cool outside" or "Set your Honeywell thermostat to a specific temperature if the temperature outside drops" | Weather |
| (6) | "Circulate Air - Clear The Smoke!" or "If tado°switched to Away mode, activate automatic control for multiple zones" | Other |
| *Ambient Luminance* | | |
| (7) | "Turn hue lights on with Siri" or "Turn lights out for bed with Alexa and hue" | Voice |
| (8) | "Turn on/off your lights with one tap on your phone" or "Toggle Yeelight on/off" | Button |
| (9) | "Turn your lights on automatically as you arrive home" or "Turn off my lights when I disconnect from my house's WiFi network." | Spatial |
| (10) | "Automatically turn your lights on at sunset" or "Turn on your Hue lights when your Eight alarm goes off" | Time |
| (11) | "Turn on a light scene depends on weather condition" or "If it is cloudy outside, turn on your lights" | Weather |
| (12) | "Turn on the lights when your Amazon Alexa alarm goes off" or "Turn your lights on if your Nest Protect detects a smoke alarm emergency" | Other |
| *Security* | | |
| (13) | "Tell Google Assistant to arm your Arlo" or "Arm Blink with Google Home" | Voice |
| (14) | "Push Button to Arm Blink" or "Press a button to unlock Sesame" | Button |
| (15) | "Automatically arm your Blink when you leave home" or "Turn your camera on when you leave home" | Spatial |
| (16) | "Close your MyQ garage nightly" or "Arm Blink at Time" | Time |
| (17) | "When it starts to rain, close my garage door" | Weather |
| (18) | "Turn your WeMo switch on when motion is detected" or "if Nest Protect detects smoke, set your Withings Home to Active Monitoring" | Other |
| *Alert User* | | |
| (19) | "Emergency Blink All Lights Red [on button press]" | Button |
| (20) | "Blink Lights when you enter an area" | Spatial |
| (21) | "Blink lights for bedtime - Hue" or "When my meeting is about to start, blink an LED on my desk" | Time |
| (22) | "Blink lights when it starts snowing" or "Play a spoken notification when it starts raining outside" | Weather |
| (23) | "Blink your Hue lights when your Amazon Alexa timer hits 0" or "Blink your lights when your doorbell rings" | Other |
| *Energy Saving* | | |
| (24) | "Turn off the lights when you leave home" or "Turn off your Wi-Fi [ASUS Router] when you leave to save battery power" | Spatial |
| (25) | "Every night at 12:00 AM turn the lights off" or "Turn Lights Off at Sunrise" | Time |
| (26) | "Turn off your lights when you leave your home in an Uber" | Other |
| *Ambient Atmosphere* | | |
| (27) | Enable "Sexy Time"Hue Lights With Amazon Alexa' or "Tell Siri which color to change your Philips hue lights to' | Voice |
| (28) | "Push Button - Blinking Christmas Lights" or "Start a party! Put your lights into disco mode [on button press]" | Button |
| (29) | "Make a grand entrance with Philips Hue" or "Set a Hue Scene when you arrive home at night" | Spatial |
| (30) | "Change your Hue lights to a softer white to help you sleep" or "Make bedroom light orange at 10pm." | Time |
| (31) | "Alert me via my Hue if the wind gets dangerous" or "Show me how hot its going to be with my lights" | Weather |
| (32) | "Change the color of your Hue lights when your order is in the oven" or "Control Hue With Wemo" | Other |
| *Robot Control* | | |
| (33) | "Tell Google Assistant to turn off your bot" or "Tell Alexa to start your Roomba" | Voice |
| (34) | "Clean a specific room with the touch of a button" or "Start Roomba with the press of a button" | Button |
| (35) | "When I arrive home from work, dock Roomba" or "When I leave home, start a cleaning job" | Spatial |
| (36) | "Turn on your bot at sunset" or "Schedule cleaning at a certain time everday" | Time |
| (37) | "Park Automower when it Rains" or "Park Automower when wind speed rises above high wind" | Weather |
| (38) | "Start your Roomba after each Litter-Robot cycle" or "When I answer a call, pause Roomba" | Other |
| *Control Hub* | | |
| (39) | "Tell Alexa to start a Harmony activity" | Voice |
| (40) | "Turn on a SmartThings device with one tap" or "Start Tahoma Scenario with Button" | Button |
| (41) | "When I get close to the home in location, run a scene in domovea" or "When you get home, a Connexoon scenario will be launched" | Spatial |
| (42) | "Turn Off Harmony Activity at given time and days of the week" or "Launch a Connexoon scenario/mode every day at 6:30PM" | Time |
| (43) | "When the UV index rises too high, launch a Connexoon scenario" or "Current weather change? Activate a scene on Smart Life" | Weather |
| (44) | "Start Harmony activity when WeMo is on" or "Amazon Echo Alarm starts your Harmony Activity" | Other |
| *Gardening* | | |
| (45) | "Tell Google Assistant to start watering" | Voice |
| (46) | "Start watering a zone at a certain time." | Time |
| (47) | "Delay my Rachio Sprinkler cycle when it rains." or "Stop watering if Netatmo detects strong winds" | Weather |
| *Outlet Control* | | |
| (48) | "Google assistant Two Wemo smartplugs off" | Voice |
| (49) | "Turn off your outlet with Button widget" | Button |
| (50) | "Turn on TP-Link/Kasa Outlet when Entering Area" | Spatial |
| (51) | "Turn on your outlet at sunset" | Time |
| (52) | "Turn On Device when Temp Drops Below Specified" | Weather |
| (53) | "Turn on WEMO Outlet When a Specific Blink Camera Detects Motion " | Other |
| *Other Appliances* | | |
| (54) | "Say "Alexa trigger coffee"to start the wemo coffee pot" or "Control your TV with Alexa" | Voice |
| (55) | "Prioritize a device with one click" or "Turn on the coffee machine" | Button |
| (56) | "Turn off your Wi-Fi when you leave to save battery power" or "Turn on air purifier when you are in" | Spatial |
| (57) | "Set your air purifier to Sleep Mode at night" or "Turn off your Wi-Fi for family time" | Time |
| (58) | "HOT water [when outsite temperature is lower than 20°C]" | Weather |
| (59) | "Turn on TV with Echo SmartThings & Harmony " or "Turn on your TV with a quick message to @IFTTT in Telegram" | Other |
| *Other* | | |
| (60) | "Turn On Hunter Douglas PowerView Schedules with Amazon Echo" or "Trigger Wink Shortcut from Amazon Echo" | Voice |
| (61) | "Press a button to beep a Wireless Tag on your keys" | Button |
| (62) | "Launch a scenario when I leave my home" | Spatial |
| (63) | "At sunset, activate a Wink shortcut" or "Feed my fish every day at 10am with littleBits Remote Pet Feeder" | Time |
| (64) | "Change indoor unit 1 when outside temp rises above" | Weather |
| (65) | "Turn off Wemo Switch with Temperature from Ecobee" or "If I disconnect from my WiFi, then enable a quick action" | Other |